\documentclass[10pt,letterpaper]{article}
\usepackage[top=0.85in,left=2.75in,footskip=0.75in]{geometry}

\usepackage[utf8]{inputenc}

\usepackage{textcomp}

\usepackage{fixltx2e}

\usepackage{amsmath,amssymb}

\usepackage{cite}

\usepackage{nameref,hyperref}

\usepackage[right]{lineno}

\usepackage{microtype}
\DisableLigatures[f]{encoding = *, family = * }

\usepackage{rotating}
\usepackage{color}
\usepackage{caption}
\usepackage{subcaption}

\raggedright
\usepackage[aboveskip=1pt,labelfont=bf,labelsep=period,justification=raggedright,singlelinecheck=off]{caption}

\makeatletter
\renewcommand{\@biblabel}[1]{\quad#1.}
\makeatother

\date{}

\usepackage{fancyhdr,graphicx}
\fancyheadoffset[L]{2.25in}
\fancyfootoffset[L]{2.25in}
\begin{document}
\vspace*{0.35in}

\begin{flushleft}
{\Large
\textbf\newline{ Fractional Dynamics of Network Growth Constrained by aging Node Interactions}
}
\newline
\\
Hadiseh Safdari\textsuperscript{1},
Milad Zare Kamali\textsuperscript{1},
Amirhossein Shirazi\textsuperscript{1},
Moein Khalighi\textsuperscript{2},
Gholamreza Jafari\textsuperscript{1,3,4 a},
Marcel Ausloos\textsuperscript{5,6,7}
\\
\bf{1} Department of Physics, Shahid Beheshti University, G.C., Evin, Tehran 19839, Iran
\\
\bf{2} Department of Mathematics, Tarbiat Modares University, Tehran, Iran
\\
\bf{3} The Institute for Brain and Cognitive Science (IBCS), Shahid Beheshti University, G.C., Evin, Tehran 19839, Iran
\\
\bf{4} Center for Network Science, Central European University, Nador 9, 1051 Budapest, Hungary,
\\
\bf{5} GRAPES, rue de la Belle Jardiniere 483, B-4031, Angleur, Belgium
\\Email : marcel.ausloos@ulg.ac.be
\\
\bf{6} School of Management, University of Leicester,  University Road, Leicester, LE1 7RH, United Kingdom
\\Email : ma683@le.ac.uk
\\
\bf{7} eHumanities group,  Royal Netherlands Academy of Arts and Sciences,  Joan Muyskenweg 25, 1096 CJ, Amsterdam, The Netherlands
\\Email : marcel.ausloos@ehumanities.knaw.nl

%
%





* corresponding  author E-mail: gjafari@gmail.com
\end{flushleft}
\section*{Abstract}
In many social complex systems, in which agents are linked by non-linear interactions, the history of events strongly influences  the whole network dynamics. However,
a class of  "commonly accepted beliefs"  seems rarely studied. In this paper, we examine how the growth process of a (social) network is influenced
by past circumstances.
In order to tackle this cause, we simply modify the well known preferential attachment mechanism by imposing a time dependent
kernel function in the network evolution equation. This approach leads to a fractional order Barab$\acute{\textbf{a}}$si-Albert (BA)
differential equation,  generalizing the BA model.
Our results show that, with passing time,  an aging process is observed  for the network dynamics.
The aging process leads to a decay for  the node  degree values,  thereby  creating an opposing process  to the preferential
attachment mechanism. On one hand, based on the preferential attachment mechanism, nodes with a high
degree are more likely to absorb links; but, on the other hand, a node's age  has a reduced chance for new connections.
This  competitive scenario allows an increased chance for younger members to  become  a hub.
 Simulations of such a network growth with aging constraint confirm the results found from solving
 the fractional BA equation. We  also report,  as an exemplary application,   an  investigation of the
 collaboration network between Hollywood movie actors. It is undubiously shown that a decay in the dynamics
 of their collaboration rate is found,  - even including a sex difference. Such findings  suggest a widely universal application of the so generalized BA model.



\section*{Introduction}

 The history of events plays a crucial role in the dynamics of many  political, economic, social, technological, legal and environmental (PESTLE) systems, - in particular  due to psychological phenomena, but also  biological and   physical  conditions. The future choices a system
makes are heavily influenced by  its  past \cite{arad2013past}.
Although, there are many  different models to describe  a network
evolution, effects of history and age are often not considered in  such models.
Here, we study the role of the aging process on network growth dynamics,
in the context of the  Barab$\acute{a}$si- Albert (BA) model as one of the models of choice
 for describing network dynamics based on preferential attachment
\cite{Albert2,Dorogovtsev2,Albert3,Papadopoulos}.


The BA model, inducing hubs in networks, i.e. nodes  with a very large number of incoming or/and outgoing links, characterizes one of the important
features of real networks, namely, scale invariance
\cite{Jeong,Victor,Holme,Ravasz,Newman}. Despite being
very successful in describing many properties of
real networks \cite{Albert2},  this model also implies that nodes can have an unlimited number of links, - a feature which
is
 not always in line with reality
\cite{Maru,Lehmann}.
In fact, in real networks, other factors work
against  such a growth of node degrees, which the BA model fails to account for
\cite{Albert1,Medo,Lind}. For instance, in realistic networks, like
citation and scientific collaboration networks \cite{Leicht,Local}, the world-wide web network \cite{Kumar}
and network of movie actor  collaborations \cite{Watts}, there are
important phenomena such as aging \cite{Adamic,Newman2,redner,Chechkin,EPL77rdirection}, and related aspects like those found in screening for diffusion limited aggregation \cite{MDLA}, limited capacity, as in city size or population growth
\cite{scala,amaral,Levins,Scheffer,gabaix2004evolution,vitanov2009nonlinear}, or  censorship \cite{Shirazi,smallkely,peace,galambook},  in the spread of knowledge, of friendship on electronic social networks, or in geo-political realms, which are not considered  inside the BA model.

In the network of movie actors, overtime "superstars"  (equivalent to "movie network hubs") are replaced by
new ones. Indeed, superstars get  promoted for some period of time but  eventually lose their attraction (in the eyes of the media),
 and then their appearance  gradually decays. New generations of stars
replace the old generations; a  similar cycle repeats again. In
political networks \cite{Burton,Chessa}, the influence of people
  grows and eventually decreases:  those who were political class rulers and party life  controllers
are replaced by often younger people. No one experiences unbound growth in
politics. In academia, the number of co-authors of most scientists grows, but later on decays \cite{Ravasz}.
There is no need to recall further  that in the economy, an unlimited  company growth,
after reaching some large share of the market, saturates and decays (if there is e.g. no innovation or merging) \cite{rotundo2014network}.
The argument could  surely go on for the 4th type of  power network, the military  network, as discussed in the IEMP model \cite{Domhoff}.


The  number of active friends a person can have is also
limited, because  a person  can  devote only part of his/her time to each of those
friends. Since there is a limited amount of time per day/week, one cannot  realistically
have an infinite number  of friends, - whatever a friend means e.g.  on  Facebook or  on Myspace.
Those "most important"  friends loose to be truly  "old friends" because the links age, and because new "potential friends"  and subsequently strong links appear
\cite{tufekci2008can}. This is similar to the screening effect phenomenon in Fermi systems, in which the
electronic potential is reduced by the cloud of electrons around one electron \cite{flensberg},
to the capacity limiting effect of  Verhulst on population growth \cite{Verhulst1}  or the size
dependent competition-cooperation effect in \cite{caram1,CaramCaifaAusloosProtoPRE}.

Thus,  in many real systems, each node has a  limited "capacity" for joining with other nodes. However,
according to the  BA model, nodes can connect to each other without any restriction, whence their degree
can grow boundlessly. In the BA model a "hub node"    will  always remain dominant as compared to newcomers, and its role as a hub will increase with time. To  adapt the
preferential attachment mechanism to some reality, the BA-based model  ideas should pay attention to these kinds of effects.
A number of studies have been done in this area: for example Palla et. \cite{palla2007quantifying} have
found that large groups persist longer  if they are capable of dynamically altering their membership,
suggesting that an ability to change the group composition results in better adaptability. Notice that  the behavior of small
groups displays the opposite tendency: "the condition for stability is that their composition remains unchanged"   \cite{palla2007quantifying}. We would go on, suggesting not only that clusters on the network change the number of nodes and their quality, but also attitudes, types of links, should  be considered  (to be) changing.

Strategies have been proposed based on adding new terms to the BA model \cite{Albert1,Dorogovtsev,Chen,Sarshar}.
It is worth noticing that other recent studies about growing network models include the aging of nodes as a key feature \cite{Zhu,Hajra}. However, requesting
  special {\it ad hoc} behavior is not a suitable approach to introduce history effects;  often, the resulting features
do not have a closed form solution either.

In order to mimic the age of a node,  toward describing some impact on the network  growth dynamics, we impose a time dependent  kernel on the evolution equation of  the node's degree,  within the  BA preferential attachment mechanism. A specific power law temporal
kernel function has been chosen in order  to preserve the effect of old events in comparison to recent ones.
Hence, we introduce  past information through a weighting function. An event in the far
past  is supposed to have less effect on the node's dynamics,  when compared with  the same event
happening  in a near past. As a result of this kernel, the governing equation for  the node's
dynamics is turned into a fractional \cite{SAMKO,Klafter,BJWest,Lundstrom,BJWest2,Vahabi} order BA-like equation. 

\begin{figure}
\centering
   \includegraphics[width=10cm]{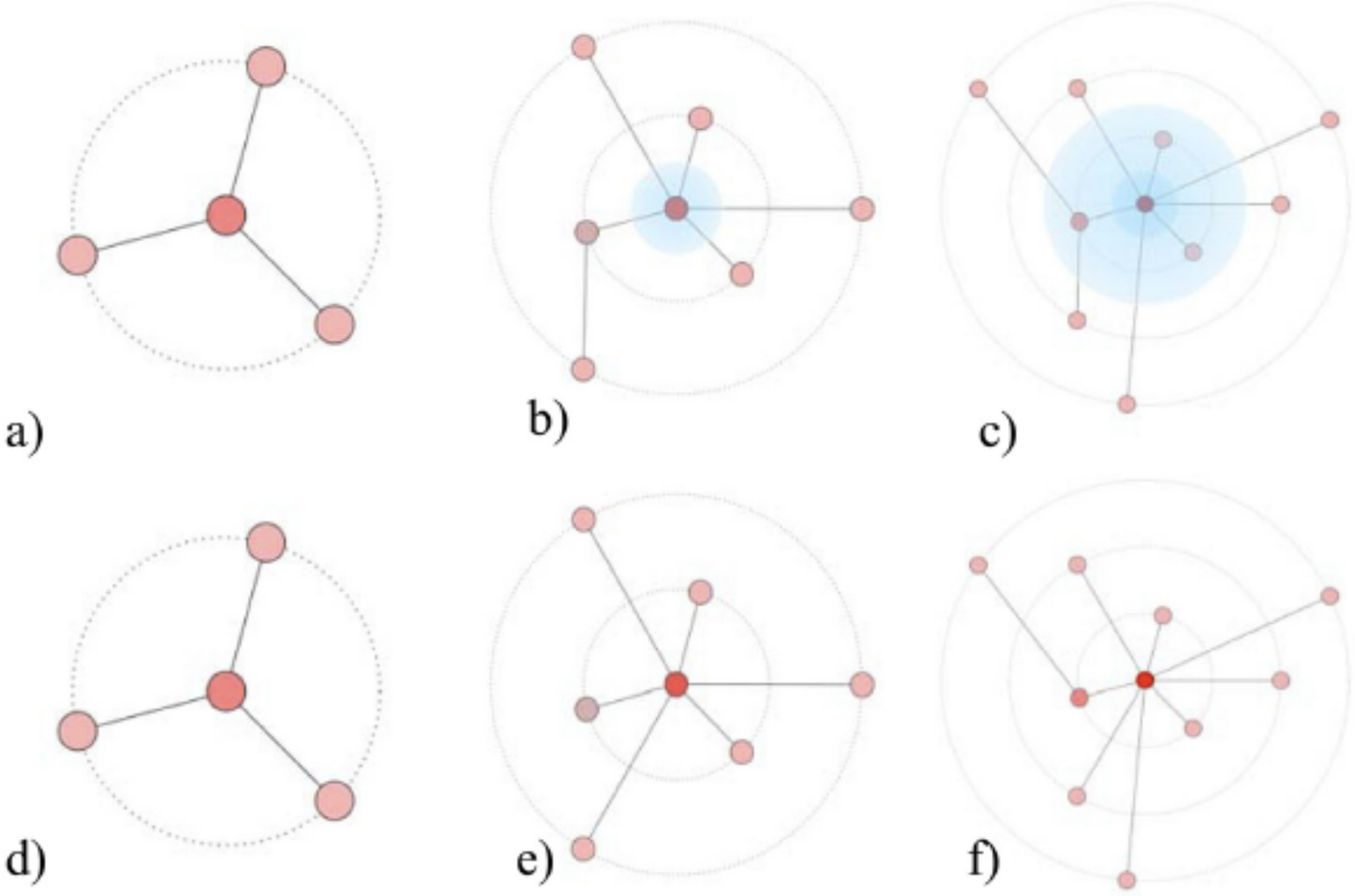}
 \caption{\small Schematic diagram representing the dynamical growth of nodes in a network,
   developing under a   preferential attachment mechanism, but also considering limitations
   imposed by an aging process and screening effects on the growth process
   (a-c);  comparisons with standard preferential attachment  are found in  (e-f).
   Red to blue color shades of nodes represent the attractiveness
   for new links. The highest attractiveness  corresponds to the  red color. The
   radial distance from the central node indicates the arrival time for a
   node. Old members, because of their age, have a lower chance of being selected. In
   (e-f), i.e. in the absence of these effects, hubs dominate the attachment process of   new links.\normalfont}
 \label{fig:schem}
\end{figure}

Figure \ref{fig:schem} schematically represents the  growth process of
a network  for which its aging process has been considered  in the dynamics.
 It is  clear from this illustration that the  limiting capacity for receiving new links and the aging of links over time
decrease the chance of high degree nodes to be attracting new links.
We have also simulated a network on which old links have less impact on a node's dynamics  when compared with newer links. In fact,
a node's effective degree  is calculated by considering the age of each link: thereafter, each node attracts new connections according to its effective degree.

Besides the numerical solution of the differential equation and analyzing the characteristics of
the simulated networks, as an actual example, we have studied the network of collaboration between Hollywood movie actors.
By following their cooperation dynamics along their career time, the aging effect on this network evolution can be observed. It is found  that the actor  network dynamics is in close agreement with the presented model,
thereby suggesting a wide application range of our model toward many other real cases.

\section*{ Methods}
\subsection*{Imposing temporal kernel on preferential attachment algorithm.}

To take into the account the aging effects, we start with the preferential attachment rule, according to which
at each time step a new node joins the system and links to $m$ already presented nodes.
Each present node can receive a new link  with a probability proportional to its degree. Beside this mechanism,
we add an aging process, which reduces the effect of each link.
Indeed the past degrees of a node are used as weight factors, which account for aging of its links \cite{West,MAINARDI,Goychuk,Metzler}. Combination of these two processes, namely, preferential attachment and aging of links, results in  a BA like integro-differential equation with a time kernel, $\kappa (t-t')$,  when integrating over the degree evolution, $dk_i(t)$. Thus, one can write,
\begin{eqnarray}
\label{kernel1}
\frac{dk_{i}(t)}{dt}=\int_{t_{0}}^{t}  dt'  \kappa (t-t')  \frac{mk_{i}(t')}{\sum_{j}^{t'} k_{j}(t')}.
\end{eqnarray}
The degrees of nodes couple to each other through
summation in the denominator over degree of all nodes up to previous time.
In case of non-aged systems, the kernel should be a Dirac delta
function, $\delta(t-t')$, which would result  in an  integer order integro-differential
equation, as  in the BA model for network growth.

Here, we choose a power law functionality for the kernel,  for describing aging process. By this choice, over time,
far past links lose their effectiveness more compared to nearly connected links.
This kernel,
guarantees the existence of scaling features as  they  contain  the intrinsic nature of most phenomena
\cite{Hansen,Mantegna,peng}. Most of the time, a distant past event should have far less effects on present ones,
when  compared with near past events;  the only exceptions are large influential events which even overshadow recent ordinary
events.  To model these exogenous causes, extra parameters  should be introduced
\cite{ZIDMA,JAS_NKVMA}.  By substituting the power law kernel in Eq.(\ref{kernel1}), one has
\begin{eqnarray}
\label{kernel2}
\frac{dk_{i}(t)}{dt}=\frac{1}{\Gamma(\alpha-1)}\int_{t_{0}}^{t} dt'
(t-t')^{\alpha-2} \frac{mk_{i}(t')}{\sum_{j}^{t'} k_{j}(t')}.
\end{eqnarray}
From fractional calculus methods,  it is apparent that the right hand side of this equation is a fractional integral of order $(\alpha-1)$:
$_{t_{0}}D_{t}^{-(\alpha-1)}$, on the interval $[t_{0},t]$
\cite{SAMKO}. Therefore, it can be obtained that
\begin{eqnarray}
\label{kernel3}
\frac{dk_{i}(t)}{dt}=_{t_{0}}D_{t}^{-(\alpha-1)} \left[ \frac{mk_{i}(t)}{\sum_{j}^{t} k_{j}(t)} \right].
\end{eqnarray}
For $\alpha=1$, the fractional operator turns to unity operator, $_{t_{0}}D_{t}^{0}\equiv 1$, hence standard BA model could be recovered.

Now applying a fractional Caputo derivative of order $(\alpha-1)$ \cite{caputo} on both sides of the above equation, we can write it in the form of a differential equation,
\begin{eqnarray}
\label{kernel4}
 _{t_{0}}^{c}D_{t}^{\alpha}\left[ k_{i}(t)\right]  = \frac{mk_{i}(t)}{\sum_{j}^{t} k_{j}(t)},
\end{eqnarray}
with each node's initial degree as $k_{i}(t_{i0}) = m$. In this way, the governing equation is a fractional order
differential equation, which guarantees the presence of aging.
More details on fractional derivatives and  relevant methods for numerical solutions can be found in the appendix.
The discrete form of the fractional order differential equation (\ref{kernel4}) becomes,
\begin{eqnarray}
\label{disk}
k_{n}=k_{0}+h^{\alpha} \Sigma_{j=0}^{n-1} b_{n-j-1} \frac{m k_{j}}{\sum_{j}^{t} k_{j}(t)}.
\label{eq_num}
\end{eqnarray}
Here, the $b_{n}$'s are time dependent coefficients,  which measure the
aging effect.  They are further outlined
 in the appendix.

These $b_n$ factors indicate the contribution of the previous degree of the
$i$-th node toward its present value. With  passing time (increasing
$n$), each  $b_{n}$ becomes smaller. In other words, over the time, the old links of a node
  lose their effect on  the node growth process  due to the $b_{n}$ coefficients  limiting form.
Therefore, the effective degree of a node, Eq. (\ref{eq_num}), will decrease.
Notice that for $\alpha=1$,  each $b_{n}$ converges
toward unity:  one gets back to the standard BA model in which old degrees
all have the same weight. 

\section*{Results}
\begin{figure}\centering
    \includegraphics[width=13cm,height=7.5cm,angle=0]{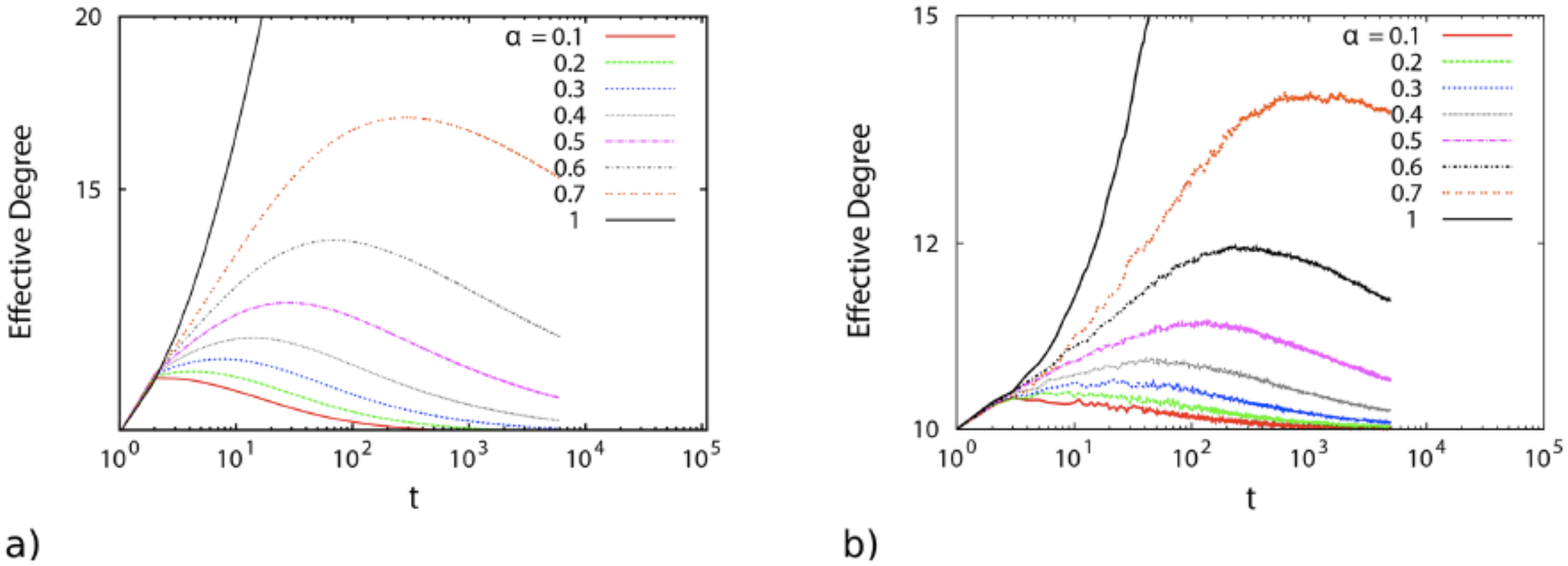} 
\caption{ (a) is the numerical solution of Eq.
  (\ref{eq_num}) for the mean degree $k$;  (b) is the corresponding mean degree $k$  of nodes from
   network simulation, when its  growth dynamics is affected by an aging
  process. It is clear from such graphs that $k(t)$ behaves differently from the
  unlimited growth predicted by the BA model ($\alpha=1$). Due to the aging process $k(t)$
  reaches a peak and declines gradually afterwards. Such simulations have
  been performed for $5000$ time steps, with as initial condition $m_{0}=11$ nodes and
  every new node connecting to $m=10$ earlier nodes.}
\label{simulation}
\end{figure}
\begin{figure} 
  \centering
    \includegraphics[width=8.5cm,height=5.0cm,angle=0]{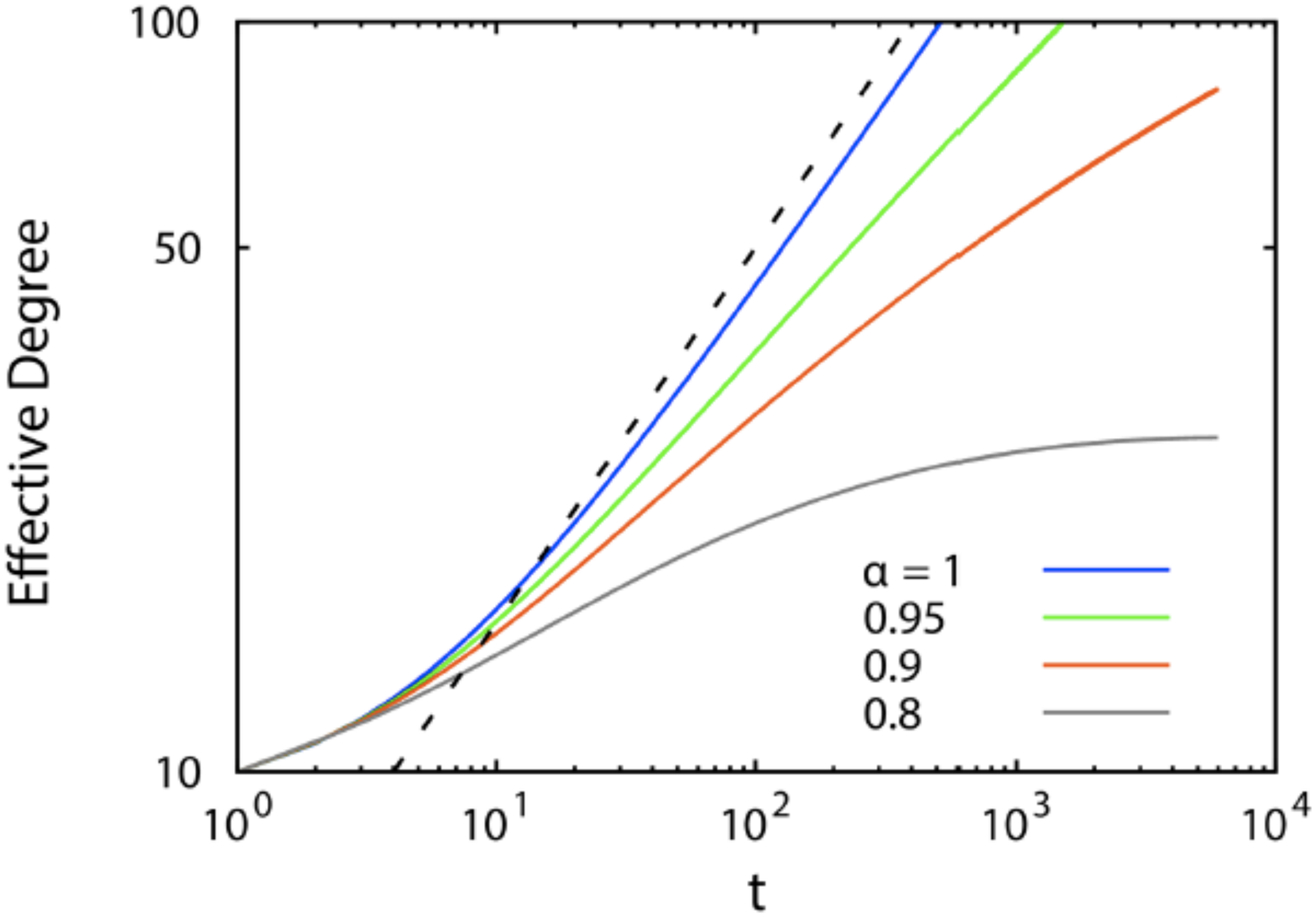}
\caption{Comparison between the numerical solution of Eq. (\ref{eq_num}) (solid lines) for exponents close to unity
and the solution of  the standard BA model (dashed line). For long times, the solution of the fractional order equation of network growth dynamics for $\alpha=1$
tends to the standard BA model.} 
  \label{fig:asympt}
\end{figure}
 Numerically solving the equation system, Eq.(\ref{disk}), we find the
 results confirming the effect of aging on the network evolution. Fig. \ref{fig:num1}
illustrates the time dependence of the  effective degree of a node at time $t$ for various  $\alpha$ values.
 It is obvious that for all values of $\alpha < 1$, the effective degree of a node $i$
 increases at first, reaches a  maximum value, and then starts to decay.
\begin{figure*}[t]
  \centering
    \includegraphics[width=13cm,height=7.5cm,angle=0]{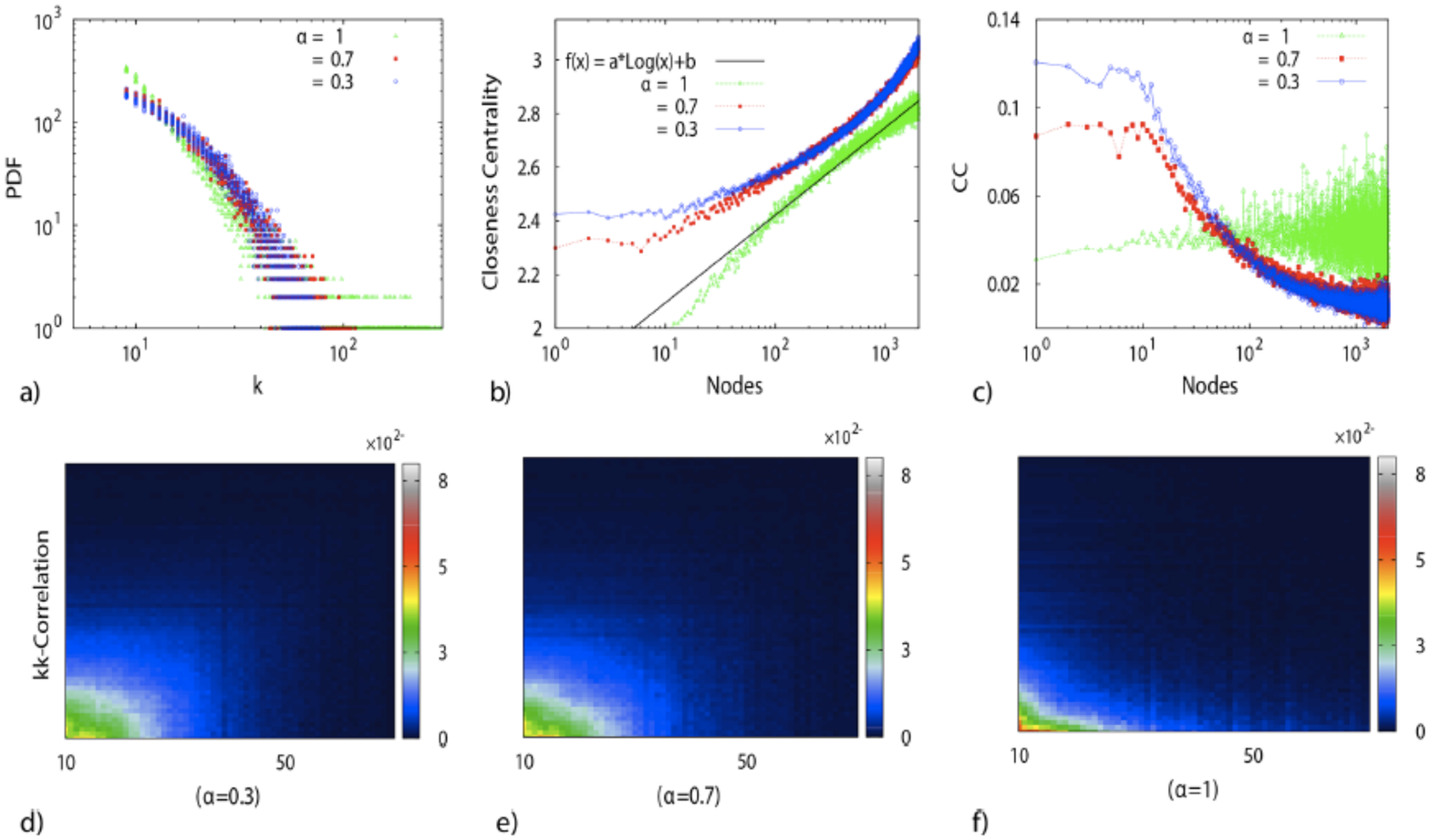}
\caption{ Panel (a) shows the probability distribution function
  for different orders of the fractional derivative. While for $\alpha=1$, i.e. the
  BA model, the expected power law behavior is found, for fractional
  orders less than unity, there is noticeable deviation from a power law. This could be caused by
  two competitive processes, namely a preferential attachment mechanism
  and a screening effect. Panel (b) is the network closeness centrality measure
  averaged over all nodes; the horizontal axis is the label of nodes according to their
  arrival time. The comparison for three exponents $\alpha$  point to different behaviors.
   For high values of $\alpha$, this measure has
  high amounts. Elder nodes have short   paths to others. The black solid line shows the fitted
  curve for $\alpha=1$.
  Panel (c) indicates the network clustering coefficient  averaged over
  all nodes for three $\alpha$ exponents. The horizontal axis also gives
 the label of nodes according to their arrival time. Older nodes have more communities around themselves.
  Panels (d-f) show the degree-degree
  correlation averaged over all nodes for the three $\alpha$ exponents.
  The tendency to bind with similar members is shown to have been
  increased by including the effects of past events. The
  inclination toward linking to previously powerful nodes is reduced. This results from the competition between new
  generations.\normalfont}
  \label{fig:image2}
\end{figure*}
In the preferential attachment mechanism, a node absorbs new links proportional to its degree.
However, here, two different processes  control the growth dynamics.
A node with a high degree  has a high chance of receiving new links. However,
these nodes usually have more aged links, which makes their effective degree smaller and in turn
the probability of new link attachments to such a node decreases.
 Figure \ref{fig:asympt} shows the convergence of the asymptotic behavior of  the node's
degree in an aged network for $\alpha=1$ with respect to the corresponding result of the standard BA model.

So far, we have seen the effects of aging on the node's
evolution in a network. In order to see the effects of aging on other usual characteristics of a network, we have simulated an "aged network".
To do this simulation, we start with an initially fully connected network with $m+1$ nodes. At each time step
a node with $m$ links is added to the network. We record the link's age, i.e. the time when a link is added to each of the $m$ nodes.
New links are connected to already presented nodes, which are selected according to their effective degree. This effective degree is found
depending on the age of each link at that time step as,
\begin{equation}
 k_{effective}(t)_i=\Sigma_{t'=1}^{t-1}\left(b_{(t')} * dk_{i}(t') \right).
\end{equation}

%

As should be expected,  for  the simulated
networks, the effective degree  experiences a growth   followed by  some  decay, as shown in Figure 2 (b). 
This observed behavior is in agreement with Fig. 2 (a), for the numerical solution of  the fractional order differential equation, Eq.(\ref{disk}).

In  Fig.  \ref{fig:image2}, we  also show  some statistical proprieties of the simulated networks.

 \begin{itemize}
\item Recall that the degree distribution is one of the major characteristics of a network. For  the
preferential attachment model,  the degree distribution follows a power law form \cite{Albert2}.
We have shown here above that in presence of aging,  deviation from  a power law occurs, see Fig.
 \ref{fig:image2} (a). This deviation is of course understood as  resulting from the competition
between the aging process
 and the preferential attachment mechanism. According
to the preferential attachment mechanism, nodes with a high degree  are prone to absorb new
links; however,  in the present case, getting older reduces the probability of
nodes with old links to be selected \cite{Lehmann}, - since the aging
reduces the effective degree of the  nodes.  This reduces the growth rate of
older nodes, whence giving nodes with a smaller degree a higher chance to
receive new links.

\item
The average shortest path to all nodes in a network \cite{Freeman} is
known as the closeness centrality measure. It can be seen from
Fig. \ref{fig:image2} (b) that the closeness centrality for the nodes in "our"
 aged networks    is  increased to that found  for the BA model. Although the centrality of older
nodes in aged networks is reduced  as compared to the BA model, their centrality is still much larger as compared to younger nodes in the same network.

\item
The clustering coefficient, Fig. \ref{fig:image2} (c), is an indicator that
displays the tendency of node's  direct neighbors to connect  with  each other
\cite{Watts}. Practically, senior members are associated with communities which
have a high degree of connectivity, as opposed to recently added members.
In other words, although  the degrees of old nodes  are heavily influenced by the aging process, these old nodes still
preserve their strategic position among others. Moreover, it can be observed that  old nodes in
a network with aging process have a higher clustering coefficient,  when
compared  with those in the same group in the BA model.

\item
One significant consequence of   aging is that new node members
have the opportunity to develop into a hub, as opposed to BA model where
only early members have a chance of becoming a hub. Hence, the tendency of
nodes to connect to their similar nodes, so called assortativity
\cite{Newman4}  should increase in networks with aged links. This can be seen in the degree-degree
correlation palette in Figs. \ref{fig:image2} (d-f). Despite the fact
that in the BA model,  the largest number of connections are located on hubs  and  old
nodes, in aged networks, a large number of connections can be found between
 nodes with the same degree.
\end{itemize} 

\subsection*{Collaboration Network of Oscar winners }
\label{section4}
As a case study we look at  the collaboration network of Hollywood movie
actors, which we extract  from the Internet
Movie Database available at $www.imdb.com$. We exclude TV movies and
TV series from the data, and build the network  strictly based on movies for theaters. For
each year,  we create a network with the condition that two nodes are
connected only if the corresponding actors/actresses have at least been cast
in a movie   together. Thus, we can discuss the network each year, and we can analyze the dynamics of this network  as a function of  time.

\begin{figure}[t]
\centering
   \includegraphics[width=13cm,height=7.5cm,angle=0]{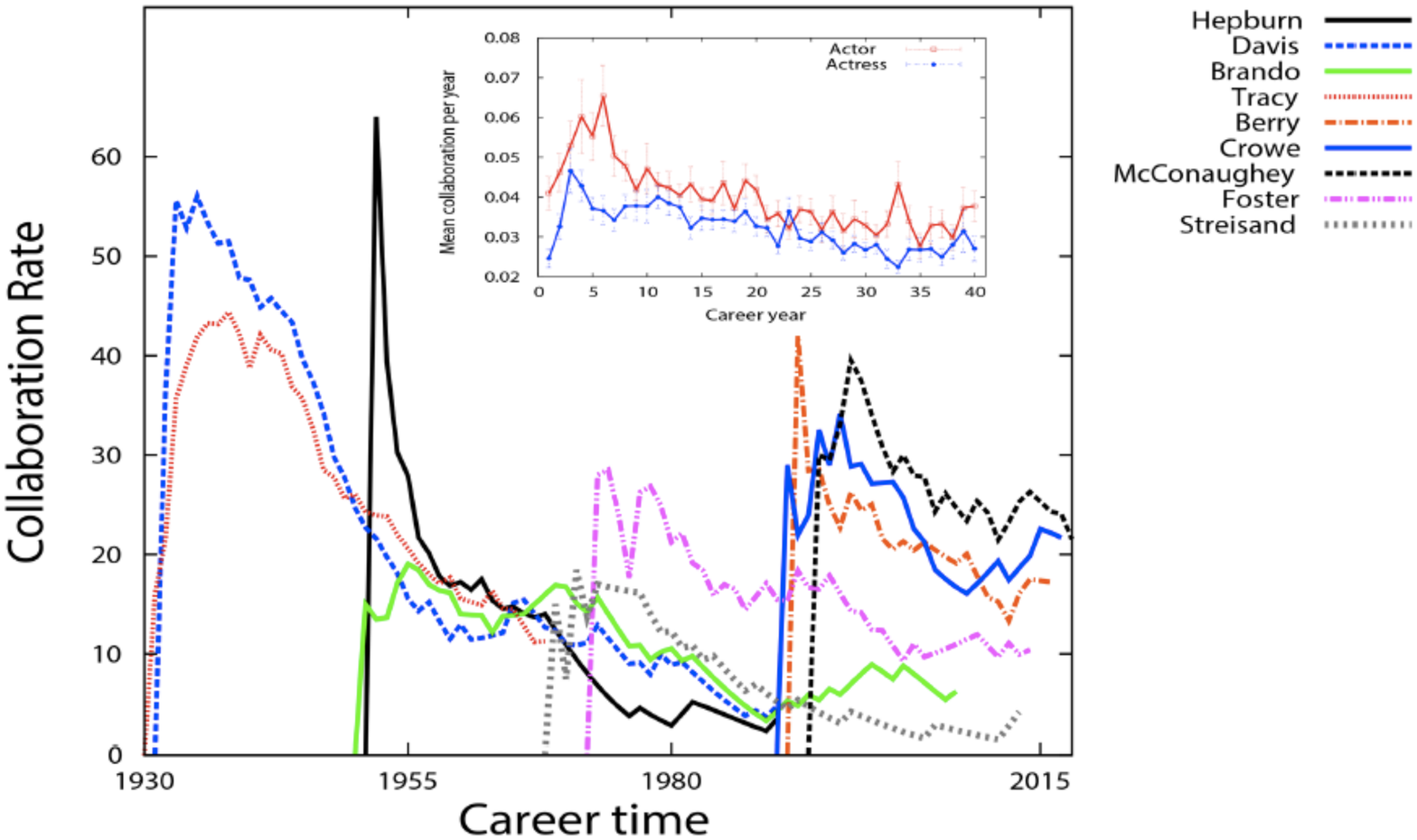}
\caption{  A sample of the total collaboration per year for Oscar
        winner actors. The "collaboration rate" is the number of staring
        actors/actresses an actor/actress has played with in a movie each year. It is
        clear that each actor/actress experiences a peak followed by a long
        decay in their total collaboration as they age
        professionally.   This peak {\it   results from the   large  number of movies a
        successful actor/actress has played at the beginning of  his/her career before   making a
        name for him/herself}. Inlet: the average number of collaborations
        per year for the 75 Oscar winner actors and 74 {\it Oscar winner} actresses. The period, where the actor/actress is making his/her
        name before the pick, is more visible here. The numbers are normalized by
        the sum of all collaborations during the {\it career}  before calculating the average. The rate of
collaboration in both curves rapidly increases;
after reaching a maximum value, it  smoothly decays. }
\label{plot_ave_actors}
\end{figure}

In  such a collaboration network, we take a closer look at the collaboration
dynamics of Oscar winners actors/actresses. We measure the collaborations of each actor/actress in each year. Fig. \ref{plot_ave_actors}
shows this history for a number of Oscar winners(a
gaussian kernel with $\sigma=5$ is also applied to data).
It can be seen that almost all the actors have the same career pattern. At the start of their acting career, the
collaboration rate increases, up to a maximum. After that, the collaboration rate gradually decreases. Despite some exceptions,
the general behavior of the collaboration follows the same pattern.

The average collaboration rate of the Oscar winning $75$ actors and
$74$ actresses present a similar pattern to the aging network model,
(we averaged the history of all actors, starting from their first year career; 
see inset in Fig. \ref{plot_ave_actors}). Although the general
behavior is similar for both curves, it is noticeable that the mean value for
the number of collaborations for actresses reaches its maximum sooner
than  for actors. It is as if the actress dynamics equation has a
smaller order of differentiation, $\alpha$, as compared to that
for actors collaboration. In other words, it is demonstrated
that aging process has a more severe impact on the actresses.

Moreover, if we divide the history of cinema into two equal parts,
there is  no noticeable difference in dynamical behavior;   the mean
 rate of collaboration in the first and the second half is the same,
indicating some reasonable validity (a "stationary system" or  "time independence"), thus giving an {\it a posteriori}
argument for the presented model. 


Therefore, we can conclude that the above results are in remarkable
agreement with findings for the numerical analysis and the  subsequently reported
simulations in the above sections, in particular in Fig. \ref{simulation}.

\section*{Discussion}

Correlations between events are  intrinsic  features of  many complex systems.
In other words, what happened before  an event will impact the current situation:  in many cases, the  history of
a real system cannot be ignored in  order to describe its  dynamics.

There are models which attempt to include history in a network evolution  through a deletion process.
Most of these models do so by adding a term to the governing differential equation.
Since these additional terms
are not unique, there is no clear method to prefer one over the other. Furthermore, such models  seem to emphasize that the evolution is much more controlled by the physical environment than by endogenous (social) features. Moreover,  often these additional
terms do not lead to an analytically tractable solution.

In an attempt to describe these deletion processes on current events
and the growth dynamics of networks, we used a kernel function as an
averaged measure of the past events within the standard differential equation of the BA model.
 This approach leads to a fractional order BA differential equation.  According to our findings,
 the presence of fractional order operators imposes a limit on nodes growth, which is in line with the observed behavior of many
 real system. This suggests that fractional calculus can be a suitable candidate for introducing history and aging in a system.

Whence, by simply generalizing the governing equation in the BA model to an
equation with fractional order derivative, a more realistic dynamics
for the systems is achieved.  As it can be known,  in many real systems,  such
as  networks of scientific papers or co-authors, friendship networks, political life, etc.,
including as discussed movie actors co-staring, the degree of nodes progresses for a certain
period of time and then smoothly decays thereafter. This change in topology of the system causes a
 realistic redistribution of the   active importance of the network members.
\section*{Acknowledgments}

\nolinenumbers

%
%
%

\subsection*{Appendix}
\subsubsection*{Fractional Derivative.}
\label{Appendix1}

For a continuous function $f$ on the $[a,b]$ interval,  the left Caputo
derivative of order $\alpha$ is defined as follow \cite{torres}:
\begin{eqnarray}
\label{eqf-1}
^{c}_{a}D_{t}^{\alpha}\left[f(t) \right]=
\frac{1}{\Gamma(n-\alpha)}\int_{a}^{t}(t-\xi)^{n-\alpha-1}(\frac{d}{d\xi})^{n}f(\xi)d\xi,
\end{eqnarray}
where $n$ is the smallest integer greater than or equal to $\alpha$,
$n=\left[ \alpha \right]+1$. The Caputo derivative has the advantage that
in solving fractional differential equations (FDE), it uses integer order
boundary or initial conditions.

In attempting to deploy a fractional calculus in the BA model of growing
networks, we start  with the dynamics equation,
\begin{eqnarray}
\label{frac_ba}
\begin{array}{lr}
^{c} _{t_{0}} D_{t}^{\alpha} k_{i}(t) = m \frac{k_{i}(t)}{\sum_{j}^{t} k_{j}(t)},   \\
k_{i}(t_{i0}) = m,
\end{array}
\end{eqnarray}
where $0< \alpha \leq 1$. For $\alpha=1$, the above equation becomes
the well-known dynamics equation in the BA model. One has a system of fractional order differential equations coupled by the summation in denominator.
From here the problem becomes an initial value problem for FDE,
\begin{eqnarray}
\begin{array}{lr}
\label{fde}
^{c} _{t_{0}} D_{t}^{\alpha} y(t) = f(t,y(t)),   \\
y(t_0) = y_{0},
\end{array}
\end{eqnarray}
which can be solved numerically by the predictor-corrector algorithm
\cite{Diethelm,Garrappa}. Hence, one can reformulate  it to an
equivalent Volterra integral equation in the form,
\begin{eqnarray}
\label{Voltera}
y(t)=y_{0}+\frac{1}{\Gamma({\alpha})}\int_{t_{0}}^{t}(t-s)^{(\alpha-1)}f(s,y(s))ds.
\end{eqnarray}

To deal with this integral, one can use the product rectangle method, which
divides the domain into $n$ fragments, $t_{j}=t_{0}+jh$, with equal
spaces $h$. The right hand side function, in terms of  such a numerical
approximation $y_{j}$ for $y(t_{j})$, is denoted by
$f(t_{j},y_{j})$. Finally,   the discrete form  reads
\begin{eqnarray}
\label{disy}
y_{n}=y_{0}+h^{\alpha} \Sigma_{j=0}^{n-1} b_{n-j-1}f_{j},
\end{eqnarray}
 with $b_{n}$ coefficients, which are obtained by using the product rectangle rule in the equispaced case \cite{Diethelm,Garrappa}, 
\begin{eqnarray}
\label{bcoeff}
b_{n}= \frac{(n+1)^{\alpha}-(n)^{\alpha}}{\Gamma(\alpha+1)}.
\end{eqnarray}
 These coefficients determine contribution of past degrees on the current degree. Smaller exponent $\alpha$ means smaller $b_n$ coefficients, which
leads to faster decay in degree of nodes. 

\bigskip

Supporting Information
S1 Text. Collecting data. In order to collect data for Fig 5 we wrote a web crawler to scan
Imdb. For each Oscar winner actor/ actress whose name we obtained from Wikipedia, we
scanned his/her Imdb page. Each person in Imdb has a page with a list of movies they have participated
in with a corresponding release date. For each movie in a person page we scanned the
movie page which contains the lead actors in that movie. Since each person has a unique id in
Imdb it is relatively simple to check how many people a person has collaborated with in each
year. Since movies are multi year projects, we apply a Gaussian kernel to the raw data before
plotting individuals in Fig 5.
\end{document}